\documentclass[%
 reprint,
superscriptaddress,
 amsmath,amssymb,
 aps,
 prd,
floatfix,
]{revtex4-2}

\usepackage{xcolor}
\usepackage{graphicx}
\usepackage{dcolumn}
\usepackage{bm}
\usepackage[pdftex]{hyperref} 

\usepackage{multirow}
\usepackage{lineno}

\newcommand{\xContained}{$10$~cm}
\newcommand{\numberHitsTrackCut}{30}
\newcommand{\vertexFlashDistanceCut}{$150$~cm}
\newcommand{\vertexTrackDistanceCut}{$5$~cm}
\newcommand{\xFiducial}{$12$~cm}
\newcommand{\yFiducial}{$35$~cm}
\newcommand{\zFiducialStart}{$25$~cm}
\newcommand{\zFiducialEnd}{$85$~cm}
\newcommand{\umu}{$\lvert U_{\mu4}\rvert^2$}
\newcommand{\umusq}{\lvert U_{\mu4}\rvert^2}
\begin{document}


\begin{flushright}
     FERMILAB-PUB-19-581-ND 
\end{flushright}

\title{Search for heavy neutral leptons decaying into muon-pion pairs in the MicroBooNE detector}

\newcommand{\Bern}{Universit{\"a}t Bern, Bern CH-3012, Switzerland}
\newcommand{\BNL}{Brookhaven National Laboratory (BNL), Upton, NY, 11973, USA}
\newcommand{\UCSB}{University of California, Santa Barbara, CA, 93106, USA}
\newcommand{\Cambridge}{University of Cambridge, Cambridge CB3 0HE, United Kingdom}
\newcommand{\StKates}{St. Catherine University, Saint Paul, MN 55105, USA}
\newcommand{\Chicago}{University of Chicago, Chicago, IL, 60637, USA}
\newcommand{\Cincinnati}{University of Cincinnati, Cincinnati, OH, 45221, USA}
\newcommand{\CSU}{Colorado State University, Fort Collins, CO, 80523, USA}
\newcommand{\Columbia}{Columbia University, New York, NY, 10027, USA}
\newcommand{\Davidson}{Davidson College, Davidson, NC, 28035, USA}
\newcommand{\FNAL}{Fermi National Accelerator Laboratory (FNAL), Batavia, IL 60510, USA}
\newcommand{\Granada}{Universidad de Granada, E-18071, Granada, Spain}
\newcommand{\Harvard}{Harvard University, Cambridge, MA 02138, USA}
\newcommand{\IIT}{Illinois Institute of Technology (IIT), Chicago, IL 60616, USA}
\newcommand{\KSU}{Kansas State University (KSU), Manhattan, KS, 66506, USA}
\newcommand{\Lancaster}{Lancaster University, Lancaster LA1 4YW, United Kingdom}
\newcommand{\LANL}{Los Alamos National Laboratory (LANL), Los Alamos, NM, 87545, USA}
\newcommand{\Manchester}{The University of Manchester, Manchester M13 9PL, United Kingdom}
\newcommand{\MIT}{Massachusetts Institute of Technology (MIT), Cambridge, MA, 02139, USA}
\newcommand{\Michigan}{University of Michigan, Ann Arbor, MI, 48109, USA}
\newcommand{\Minnesota}{University of Minnesota, Minneapolis, MN, 55455, USA}
\newcommand{\NMSU}{New Mexico State University (NMSU), Las Cruces, NM, 88003, USA}
\newcommand{\Otterbein}{Otterbein University, Westerville, OH, 43081, USA}
\newcommand{\Oxford}{University of Oxford, Oxford OX1 3RH, United Kingdom}
\newcommand{\PNNL}{Pacific Northwest National Laboratory (PNNL), Richland, WA, 99352, USA}
\newcommand{\Pitt}{University of Pittsburgh, Pittsburgh, PA, 15260, USA}
\newcommand{\Rutgers}{Rutgers University, New Brunswick, NJ, 08901, USA}
\newcommand{\StMarys}{Saint Mary's University of Minnesota, Winona, MN, 55987, USA}
\newcommand{\SLAC}{SLAC National Accelerator Laboratory, Menlo Park, CA, 94025, USA}
\newcommand{\SDSMT}{South Dakota School of Mines and Technology (SDSMT), Rapid City, SD, 57701, USA}
\newcommand{\Syracuse}{Syracuse University, Syracuse, NY, 13244, USA}
\newcommand{\TelAviv}{Tel Aviv University, Tel Aviv, Israel, 69978}
\newcommand{\Tennessee}{University of Tennessee, Knoxville, TN, 37996, USA}
\newcommand{\UTA}{University of Texas, Arlington, TX, 76019, USA}
\newcommand{\Tufts}{Tufts University, Medford, MA, 02155, USA}
\newcommand{\VTech}{Center for Neutrino Physics, Virginia Tech, Blacksburg, VA, 24061, USA}
\newcommand{\Warwick}{University of Warwick, Coventry CV4 7AL, United Kingdom}
\newcommand{\Yale}{Wright Laboratory, Department of Physics, Yale University, New Haven, CT, 06520, USA}

\affiliation{\Bern}
\affiliation{\BNL}
\affiliation{\UCSB}
\affiliation{\Cambridge}
\affiliation{\StKates}
\affiliation{\Chicago}
\affiliation{\Cincinnati}
\affiliation{\CSU}
\affiliation{\Columbia}
\affiliation{\Davidson}
\affiliation{\FNAL}
\affiliation{\Granada}
\affiliation{\Harvard}
\affiliation{\IIT}
\affiliation{\KSU}
\affiliation{\Lancaster}
\affiliation{\LANL}
\affiliation{\Manchester}
\affiliation{\MIT}
\affiliation{\Michigan}
\affiliation{\Minnesota}
\affiliation{\NMSU}
\affiliation{\Otterbein}
\affiliation{\Oxford}
\affiliation{\PNNL}
\affiliation{\Pitt}
\affiliation{\Rutgers}
\affiliation{\StMarys}
\affiliation{\SLAC}
\affiliation{\SDSMT}
\affiliation{\Syracuse}
\affiliation{\TelAviv}
\affiliation{\Tennessee}
\affiliation{\UTA}
\affiliation{\Tufts}
\affiliation{\VTech}
\affiliation{\Warwick}
\affiliation{\Yale}

\author{P.~Abratenko} \affiliation{\Tufts} 
\author{M.~Alrashed} \affiliation{\KSU}
\author{R.~An} \affiliation{\IIT}
\author{J.~Anthony} \affiliation{\Cambridge}
\author{J.~Asaadi} \affiliation{\UTA}
\author{A.~Ashkenazi} \affiliation{\MIT}
\author{S.~Balasubramanian} \affiliation{\Yale}
\author{B.~Baller} \affiliation{\FNAL}
\author{C.~Barnes} \affiliation{\Michigan}
\author{G.~Barr} \affiliation{\Oxford}
% ...

 \collaboration{The MicroBooNE Collaboration} \thanks{microboone\_info@fnal.gov}\noaffiliation

\date{\today}

\begin{abstract}
We present upper limits on the production of heavy neutral leptons (HNLs) decaying to $\mu \pi$ pairs using data collected with the MicroBooNE liquid-argon time projection chamber (TPC) operating at Fermilab. This search is the first of its kind performed in a liquid-argon TPC. We use data collected in 2017 and 2018 corresponding to an exposure of $2.0 \times 10^{20}$ protons on target from the Fermilab Booster Neutrino Beam, which produces mainly muon neutrinos with an average energy of $\approx 800$~MeV. HNLs with higher mass are expected to have a longer time-of-flight to the liquid-argon TPC than Standard Model neutrinos. The data are therefore recorded with a dedicated trigger configured to detect HNL decays that occur after the neutrino spill reaches the detector. We set upper limits at the $90\%$ confidence level on the element $\umusq$ of the extended PMNS mixing matrix in the range $\umusq<(6.6$--$0.9)\times 10^{-7}$ for Dirac HNLs 
and $\umusq<(4.7$--$0.7)\times 10^{-7}$ for Majorana HNLs, 
assuming HNL masses between $260$ and $385$~MeV and $\lvert U_{e 4}\rvert^2 = \lvert U_{\tau 4}\rvert^2 = 0$. 
\end{abstract}


\maketitle


\section{Introduction}
\label{sec:introduction}
The standard model (SM) describes massless neutrinos as left-handed states. The observation of neutrino oscillations~\cite{Tanabashi:2018oca} has demonstrated, however, that neutrinos must have mass, requiring extensions of the SM, such as the neutrino minimal standard model ($\nu$MSM) \cite{Asaka:2005pn,Asaka:2005an}. The $\nu$MSM predicts additional right-handed neutral leptons that, unlike SM neutrinos, are not charged under the weak interaction and thus manifest themselves only through their mixing with SM neutrinos. According to the $\nu$MSM, a right-handed neutral lepton at the keV mass scale can provide a candidate for dark matter, while the other two right-handed leptons are expected to have masses at the GeV scale~\cite{Asaka:2005pn,Asaka:2005an}. In general, the masses of these right-handed states and their coupling to SM neutrinos are not predicted by the model, and their allowed values can thus span many orders of magnitude.
In this paper, we report results from a search for 
heavy neutral leptons (HNLs) with masses of ${\cal O}(100)$~MeV.

The MicroBooNE detector \cite{Acciarri:2016smi} began collecting data from the Booster Neutrino Beam (BNB)~\cite{AguilarArevalo:2008yp} in 2015, making it the first fully operational detector of the three liquid-argon time projection chambers comprising the Short-Baseline Neutrino Program (SBN) \cite{Machado:2019oxb}.
The SBN program will address the short-baseline anomalies observed by the MiniBooNE and LSND Collaborations~\cite{Aguilar:2001ty,Aguilar-Arevalo:2018gpe}.
One possible explanation of these anomalies is the existence of light sterile neutrinos with masses of the order of eV, 
which would lead to short-baseline neutrino flavor oscillations not expected in the SM.

In addition to the studies of the effects of eV-scale sterile neutrinos,
the energy range of the BNB allows us to extend the sensitivity of the SBN detectors to  
the production and decay of HNLs with masses of ${\cal O}(100)$~MeV.
HNLs produced by the BNB would travel along the beamline and 
could then decay in-flight to $\mu \pi$ pairs inside the MicroBooNE detector, located $463$~m downstream from the
neutrino production target.
Due to their mass, some of the HNLs
are expected to arrive late compared to the arrival of the BNB spill. To suppress background from SM neutrino
interactions, we therefore use data collected with a dedicated HNL trigger. This trigger was commissioned in 2017 and is used to search for late signatures occurring after the arrival of the SM neutrino beam spill.  
This trigger allows us to perform a search for HNLs in the mass range $260$--$385$~MeV
using data taken in 2017 and 2018 that corresponds to $2.0 \times 10^{20}$ protons on target (POT).

\section{Heavy Neutral Leptons}
\label{sec:properties}

We define the HNL in terms of its relevant parameters: its mass $m_{N}$, and the elements of the extended PMNS matrix $\lvert U_{\alpha 4}\rvert^2$ ($\alpha$ = $e$, $\mu$, $\tau$). The flavor eigenstates of the left-handed neutrinos $\nu_{\alpha}$ are written as a linear combination of the SM neutrino mass eigenstates $\nu_{i}$ ($i$ = 1, 2, 3) and the heavy neutral lepton state, $N$, in the form:
\begin{equation}
    \nu_{\alpha} = \sum_i U_{\alpha i} \nu_i + U_{\alpha 4} N.
\end{equation}
HNLs can be produced (and decay) via SM gauge interactions, with a rate suppressed by the relevant $\lvert U_{\alpha 4}\rvert^2$ element through mixing-mediated interactions with SM gauge bosons. The decay of charged kaons and pions from the BNB can thus produce a flux of HNLs, which then propagate to the MicroBooNE detector, where they are assumed to decay to SM particles. 

\begin{figure}[htbp]
\includegraphics[width=0.36\textwidth]{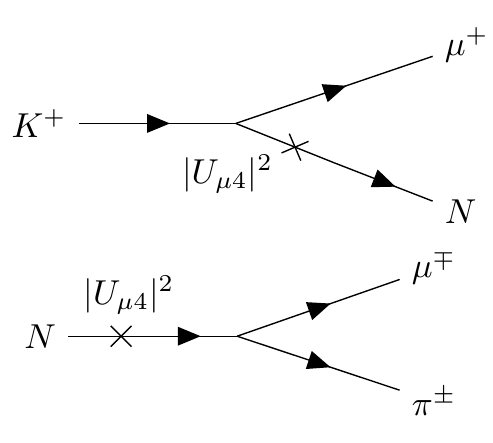}
\caption{Production of an HNL (labelled $N$) via mixing in a $K^+$ meson decay and its subsequent decay into a $\mu^{\mp} \pi^{\pm}$ pair.}
\label{fig:HNL_decay}
\end{figure}

Figure~\ref{fig:HNL_decay} shows diagrams for the production and decay channels. 
In this paper, we only consider HNL decays to $\mu\pi$ final states, where the HNLs are produced through the process $K^+ \rightarrow \mu^+ N$. 

HNL states can include both Dirac and Majorana mass terms.
Majorana HNLs would decay in equal numbers into $\mu^+\pi^-$ and $\mu^-\pi^+$ final states. Since the BNB with positive horn polarity used for this search produces predominantly neutrinos and not anti-neutrinos,
Dirac HNLs could only decay through the process $N\to \mu^-\pi^+$.

Assuming 
$\lvert U_{e 4}\rvert^2 = \lvert U_{\tau 4}\rvert^2 = 0$,
the HNL production rates and the $\mu \pi$  decay width are both proportional to $\lvert U_{\mu 4}\rvert^2$ only~\cite{Atre:2009rg}, and we therefore place limits exclusively on the \umu{} mixing matrix element.
The accessible HNL masses are given by the requirement that the decay and production be kinematically allowed, i.e.,  
$ m_K - m_{\mu}>m_N>m_{\mu} + m_{\pi}$.

The angular distributions of the decay products is given by the angle between the polarization vector of the HNL and the momentum  of the charged lepton in the HNL rest frame.
The angular distributions differ between the charge combinations ($\mu^+\pi^-$ and $\mu^-\pi^+$) but the combined distribution describing
Majorana HNLs is isotropic, and the expected rate is double the rate of Dirac HNL decays~\cite{Balantekin:2018ukw,ballett2019heavy}. 

\section{HNL Flux in the BNB}

The BNB impinges protons from the Fermilab Booster synchrotron on a beryllium target. 
The protons are delivered in a spill with a duration of $1.6$~$\mu$s and an average repetition rate of up to $5$~Hz~\cite{AguilarArevalo:2008yp}.
The proton kinetic energy of $8$~GeV limits the types of mesons produced by p-Be interactions to kaons and pions, generating a muon-neutrino beam with average energies of $800$~MeV.
This restricts the highest mass of HNL that can be produced at the BNB to $m_{K} - m_{\mu}$ for $\lvert U_{\mu 4}\rvert^2$ mediated channels. 

We calculate the HNL production rate from the BNB using the SM neutrino flux simulation~\cite{AguilarArevalo:2008yp}.
The decay kinematics of each SM neutrino parent are calculated for an HNL of mass $M_N$. Each event is then weighted by a kinematic factor to account for the effect of $M_N$ on the parent decay rate and by a geometric factor describing the probability of the HNL reaching the MicroBooNE detector. 
The geometric factor enhances the flux since the HNLs with higher mass 
are boosted into the beam direction. 

The kinematic factor suppresses HNL production at the kinematic threshold and takes into account the smaller helicity suppression due to the mass of the HNLs~\cite{Shrock:1980vy,Shrock:1980ct,Ballett:2016opr}. 

Due to the Lorentz transformation into the lab frame,
the decay probability becomes inversely proportional to HNL momentum. In contrast, the number of SM neutrino interactions is given by their interaction cross sections on argon, which rises with energy.  The HNL flux is thus expected to
be enhanced at lower momenta, leading to correspondingly longer travel times to the detector.

\section{MicroBooNE Detector}
\label{sec:detector}

The MicroBooNE detector~\cite{Acciarri:2016smi} is a liquid-argon time projection chamber (LArTPC)
situated at near-ground level at a location $463$~m downstream from the target of BNB, receiving a $93.6\%$ pure $\nu_{\mu}$ beam. The MicroBooNE TPC has an active mass of $85$~t of liquid argon, in a volume $2.6 \times 2.3 \times 10.4$~m$^3$ in the $x,y,z$ coordinates, respectively.
The MicroBooNE detector is described by a right-handed coordinate system. The $x$ axis points along the negative drift direction with the origin located at the anode plane, the $y$
axis points vertically upward with the origin at the center of the detector, and the $z$ axis
points along the direction of the beam, with the origin at the upstream edge of the detector. The polar angle is defined
with respect to the $z$ axis and the azimuthal angle $\phi$ with respect to the $y$ axis.

Neutrinos that cross the detector can interact with the argon nuclei, or, in the case of HNLs, decay to SM particles and produce secondary charged particles that ionize the argon atoms along their trajectories producing ionization electrons and scintillation light. An electric field of $273$~V/cm causes the electrons to drift towards the anode plane, requiring $2.3$~ms to drift across the width of the detector. 
The anode planes are positioned perpendicular to the electric field and comprise three planes of sense wires with a spacing of 3~mm between adjacent wires and the same spacing separating the wire planes.
Ionization electrons induce a bipolar signal when they pass through the first two planes of wires, oriented at $\pm 60^\circ$ with respect to the vertical,
before being collected on the third plane with vertically oriented wires producing a unipolar signal.

\begin{figure}[htbp]
    \includegraphics[width=0.46\textwidth]{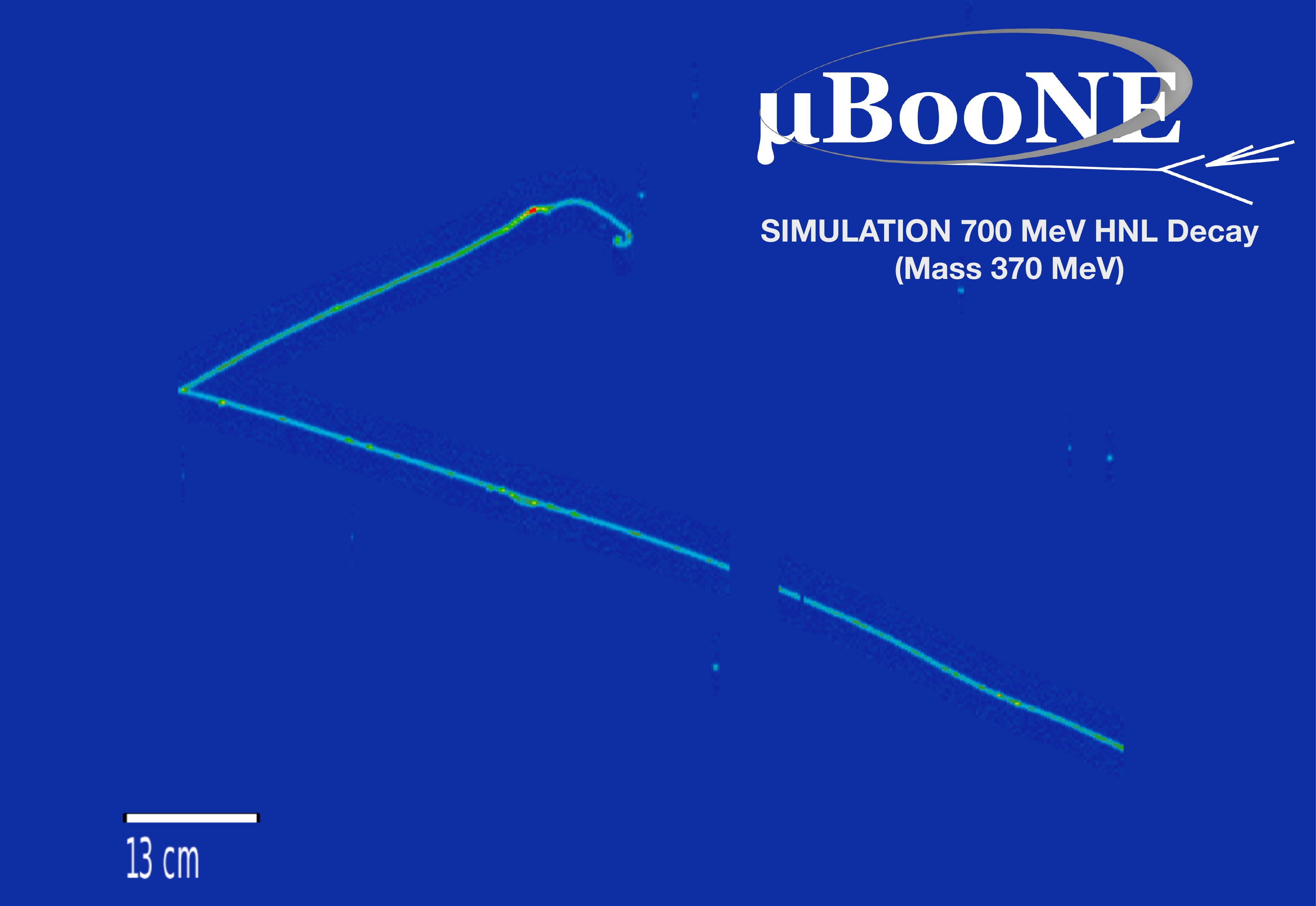}
    \caption{Display of a $\mu \pi$ decay for an HNL with a mass of $370$~MeV, showing the signals measured at the collection plane. The horizontal axis represents the wire-number, and the vertical axis represents time. Colors show the charge deposition measured on the wires. The gap in the longer track is due to a set of unresponsive wires.}
    \label{fig:evdisplay}
\end{figure}

The waveforms measured by the $8192$ wires are digitized in a $4.8$~ms readout window. The signal processing on the raw TPC waveforms includes noise filtering and deconvolution to convert wire signals into hit information~\cite{Acciarri:2017sde}.
Subsequently, individual hits corresponding to a localized 
energy deposit are extracted for each wire.
The combination of timing information and energy deposit contained in each waveform is used to create 2D projective views of the event. Fig.~\ref{fig:evdisplay} shows such a 2D view for a simulated HNL decay.
The Pandora~\cite{Marshall:2015rfa,Acciarri:2017hat} toolkit is then used to reconstruct 3D tracks (produced by muons, pions and protons) and 
showers (produced by electrons and photons) from the 2D views.

A calibration is performed to take into account all the microphysics in the detector,
including electron-ion recombination and the space charge effect (SCE)~\cite{Adams:2019ssg}.
The SCE is caused by slowly drifting ions produced by cosmic rays
that create variations in the electric fields.
This variation impacts energy deposits and track trajectories, which appear distorted, particularly near the edges of the TPC.
The individual energy deposits are corrected by the time-averaged calibration
factors obtained from data.

An array of 32 8-inch photomultiplier tubes (PMTs) with 16~ns timing resolution collects the scintillation light produced by argon ionization. We use this measurement to determine the time of the neutrino interaction and for triggering. 
Light flashes are reconstructed with a timing resolution of $100$~ns
by summing waveforms from the 32 PMTs.

\section{Triggers and Data Samples}
\label{sec:trigData}

To reduce the amount of recorded data,
online software triggers are deployed, processing the waveforms of the light collection system 
so that activities in coincidence with BNB spills are identified and stored.
SM neutrinos arrive at the MicroBooNE detector $1.5$~$\mu$s after they have been produced, while the time-of-flight of the HNLs depends on their mass and momentum.
As the HNL mass increases, an increasing fraction of HNL events would thus arrive at the detector after the end of
the BNB trigger window of $1.9$~$\mu$s.

HNLs decaying into $\mu\pi$ pairs within the BNB 
trigger window need to be discriminated from HNL-like backgrounds due to SM charged-current 
neutrino interactions that include muons, pions, and also protons in the final state. Neutrino-induced
charged-current coherent pion production, where no additional activity 
around the vertex is expected, could present a nearly irreducible background.
The only background relevant for HNL decays that occur after the end of the BNB trigger window 
are crossing cosmic-ray muons, which have a distinctly different topology from the signal.
We therefore focus on the HNLs arriving after the end of the BNB trigger window in this analysis.

\begin{figure}[htbp!]
    \includegraphics[width=0.48\textwidth]{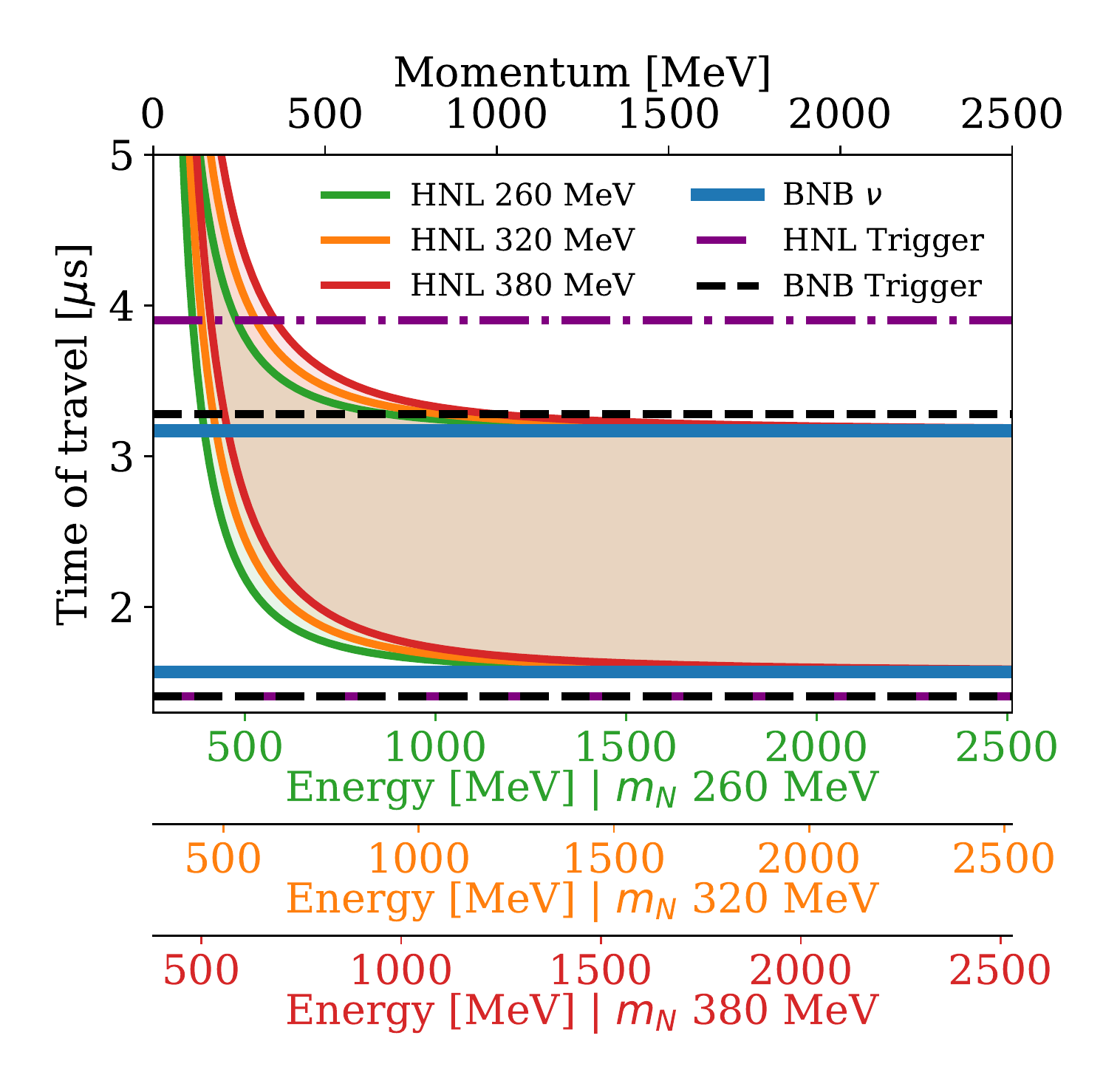}
    \caption{Time of travel from the BNB target to the MicroBooNE detector for SM active neutrinos and HNLs of different masses. Blue solid lines indicate time of travel for SM active neutrinos, produced within a beam spill of $1.6$~$\mu$s. Neutrinos are expected to arrive at any time between the two solid lines independent of their initial momentum. The dashed lines indicate the start (black at $1.4$~$\mu$s) and end time of the BNB (black) and HNL (purple) trigger. The solid lines and bands indicate the time of travel for HNLs of different mass within a spill. }
    \label{fig:triggerTimingBand}
\end{figure}

In June 2017, we introduced an HNL trigger that starts concurrently with the BNB trigger -- in coincidence with the beam spill arrival -- 
but extends the trigger window from $1.9$~$\mu$s to $2.5$~$\mu$s. 
Fig.~\ref{fig:triggerTimingBand} shows the relationship between the time-of-flight, 
the initial energy of HNLs,
and the trigger windows, illustrating that the HNL trigger retains a larger fraction of HNL decays when the HNL is heavier.
The HNL trigger requires the number of photoelectrons ($N_{pe}$) recorded by the PMTs to be $N_{pe}>10.5$, which is slightly higher than the requirement of $N_{pe}>6.5$ for the BNB trigger.
This choice optimizes signal efficiency and trigger rate.

The MicroBooNE detector is exposed to a large flux of cosmic-ray muons, traversing at a rate of $\approx 5.5$~kHz, since it is situated just below ground level at a depth of $\approx 6$~m with no significant overburden.
Events in coincidence with a BNB spill (``on-beam'') data contain
up to $\approx 20$ cosmic-ray muons within the readout window of $4.8$~ms.

For background studies, we collect data sets based on identical trigger settings as for on-beam data,
with the exception of the beam coincidence requirement. These ``off-beam''
data sets 
were taken with either the BNB or the HNL trigger requirement on the number of photoelectrons. They
contain mainly cosmic rays and no SM neutrino interactions.   

In summary, we make use of three data samples:
\begin{itemize}
    \item \textbf{On-beam HNL data}, taken in coincidence with a BNB neutrino spill. This data set requires an event to fulfil the HNL trigger condition with a veto on the BNB trigger to reject activity produced by neutrino interactions during the BNB spill. 

    \item \textbf{On-beam BNB data}, taken in coincidence with a BNB neutrino spill and fulfilling the BNB
     trigger conditions.
     
 \item \textbf{Off-beam data}, taken with identical trigger settings and in a time window of the same length as for the on-beam BNB or HNL data, but at a time when no beam spills are received at the detector. 
 
\end{itemize}
\section{Monte Carlo Samples}
\label{sec:mc}
Simulated data sets are used to evaluate the reconstruction and selection efficiency, 
to train a boosted decision tree for signal discrimination and to provide a control sample
of SM neutrino interactions for validation. 
  \begin{figure}[htbp!]
    \includegraphics[width=0.48\textwidth]{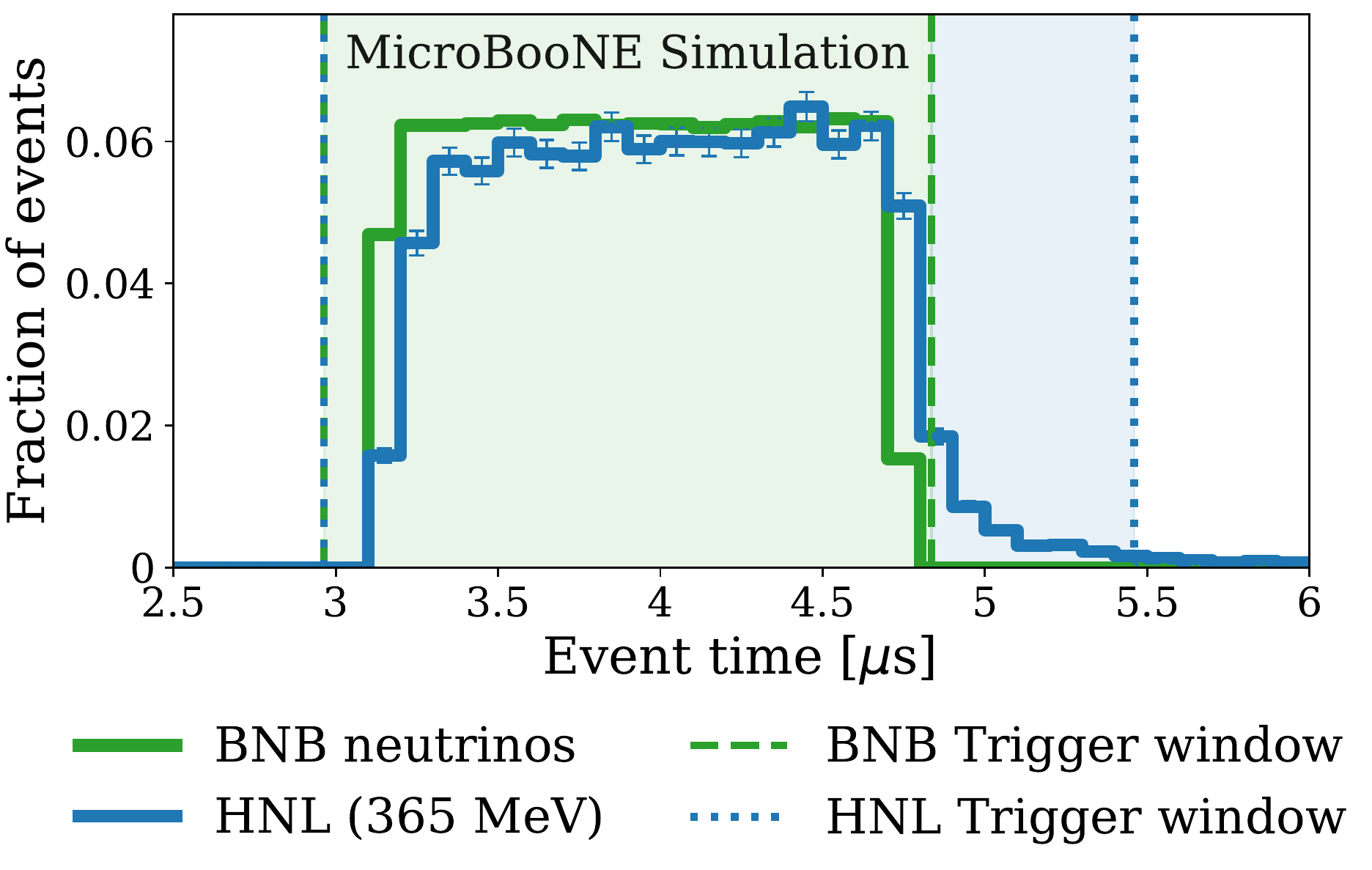}
    \caption{Timing distribution for muon-neutrinos and HNLs produced in the BNB. The HNL mass is 365 MeV.  Vertical lines indicate the start and end time of the BNB and HNL trigger windows.}
    \label{fig:triggerTiming}
\end{figure}

  HNL signal samples are simulated for ten different HNL masses.
  All HNLs are assumed to travel collinearly with the beam axis, i.e., parallel to the longitudinal $z$ axis of the MicroBooNE detector, such that $p_{x} = p_{y} = 0$. 
  The time distribution of the HNL production is assumed to be uniform within the beam spill, 
  neglecting the $\approx 19$~ns wide bunch structure, which cannot be resolved with the light reconstruction used in this analysis.
  The calculated arrival time distributions for HNLs and SM neutrinos produced in a BNB spill is shown
  in Fig.~\ref{fig:triggerTiming}. We simulate HNL decays into $\mu^{-}\pi^{+}$ and $\mu^{+}\pi^{-}$ final states with isotropic angular distributions. The $\mu^{-}\pi^{+}$ decays are reweighted to obtain the angular distribution
for Dirac HNLs.

Interactions of SM neutrinos in liquid argon, as well as in the material surrounding the detector (``dirt'' events),
are simulated within the \texttt{LarSoft}~\cite{Snider:2017wjd} framework
using the \texttt{GENIE}~\cite{Andreopoulos:2015wxa} Monte Carlo program.
For HNL signal events, cosmic rays crossing the detector are modelled by overlaying data from zero-bias off-beam data events, whereas the \texttt{CORSIKA}~\cite{Heck:1998vt} program is used to simulate cosmic rays for SM neutrino interactions.
The detector
simulation and propagation of secondary particles in liquid argon is
  simulated with \texttt{GEANT4}~\cite{Agostinelli:2002hh,Allison:2016lfl}.
   
\section{Event Selection}
\label{sec:selection}
As this analysis focuses on the HNLs arriving after the end of the BNB trigger window,
events containing signal candidates need to pass the HNL trigger condition but not the BNB trigger condition. 
Since signal candidates are expected to comprise two reconstructed tracks sharing a common vertex, we 
select vertices reconstructed by Pandora with exactly two associated 
tracks. At this stage, events can contain more than one
such HNL candidate. To reduce background, we apply further selections:

\begin{itemize}
\item {\bf Fiducial Volume.}
The reconstructed vertex associated with the HNL candidate is required to be located in a fiducial volume, defined as a cuboid contained in the active TPC volume. 
The vertex location must be greater than \xFiducial{} from the border of the active volume along the $x$~axis, \yFiducial{} along the $y$ axis, \zFiducialStart{} from the upstream edge of the $z$~axis, and \zFiducialEnd{} from the downstream edge.
HNL candidates with vertices located within a $1$~m wide gap along the $z$~axis in the
range $675<z<775$~cm are also rejected since the wires are not optimally performant in this
region~\cite{Acciarri:2017sde}.
The fiducial volume thus defined corresponds to a mass of $43$~t of liquid argon.

\item {\bf Vertex Track Distance.}
We require the starting points of the two tracks associated with the HNL candidate to lie within \vertexTrackDistanceCut{} 
from the location of the reconstructed vertex.

\item {\bf Minimum Number of Hits.}
The tracks associated with an HNL candidate are required to each have more than \numberHitsTrackCut{}
associated hits in the collection plane, which corresponds to a minimum energy deposit of $\approx 20$~MeV. 

\item {\bf Flash Distance.}
We require the distance in the $yz$ plane
between the center of the reconstructed light flash and the $yz$ projection of the vertex location to be
less than \vertexFlashDistanceCut{}.

\item {\bf Track Containment.}
The tracks associated with an HNL candidate have to be fully contained within a volume defined 
by a distance of $25$~cm from the edges of the active volume on the $y$ axis 
and \xContained{} on the $x$ and $z$ axes. 
This requirement removes tracks crossing the TPC edges that are more severely affected by the SCE. It also rejects cosmic-ray background.

\item {\bf Kinematics.}
A large fraction of cosmic-ray muons misidentified as HNL candidates are 
caused by ``broken'' tracks where the reconstruction algorithm
has split the muon track, assigning the two sections to a common vertex at the point where
the track is broken.
Such candidates have a large opening angle $\Delta\alpha$ between the two tracks 
and are thus rejected by requiring $\Delta\alpha<2.8$. We also require that the mass as determined from the momenta assigned to the tracks is $<500$ MeV. 

The momenta of the two tracks representing the $\mu\pi$ pair from the HNL decay are determined by the length of the track under the hypothesis 
that the longer particle track is from the muon and the shorter the pion. The length of the tracks associated with the HNL candidates is of the order $10$~cm and depends on the HNL mass.
The impact of the muon-pion assignment on the results is negligible.
\end{itemize}

\begin{table}[htbp!]
\caption{Number of candidates remaining after the selection requirements for 
the HNL signal with a mass of $370$~MeV and for $\umusq=1.4\times 10^{-7}$, the expected background rate derived from the off-beam data
set, and the on-beam HNL data, corresponding to $2.0 \times 10^{20}$ POT. The uncertainty on the expected background rate is given by 
the statistical uncertainty of the off-beam data.}
\begin{center}
\begin{tabular}{c|ccc}
\hline\hline
                         & HNL Signal & Background  & Data  \\ \hline
Two-track Vertex         & $100$ & 41,426                & 41,914            \\
Fiducial Volume          & $61$ & 21,501        & 21,811  \\
Vertex-track Distance    & $57$ & 16,126        & 16,339  \\
Minimum Number of Hits   & $57$ & 15,924        & 16,126 \\
Flash Requirement        & $57$ & 7,487         & 7,527 \\
Track Containment        & $47$ & 1,096         & 1,138  \\
Kinematics (mass, angle) & $45$ & $653 \pm 286$   & $669$  \\
\hline\hline
\end{tabular}
\end{center}
\label{tab:qualityCuts_efficiency}
\end{table}

After applying these selection requirements, $(45$--$50)\%$ of the HNL candidates in the HNL simulation are retained, with a corresponding
background selection efficiency of $1.6\%$.
The contribution from dirt events in the background sample are found to be negligible.
\autoref{tab:qualityCuts_efficiency} 
shows the numbers of HNL candidates for the data samples used for this analysis, corresponding to $2.0 \times 10^{20}$ POT, and for the corresponding HNL signal simulation
assuming a mass of 370 MeV and a mixing angle of $\umusq=1.4\cdot 10^{-7}$. 
The expected number of background candidates is derived from the off-beam data
by normalizing the number of time windows in the off-beam data to the number of beam spills of the on-beam data. The background expectation agrees with the data within the statistical uncertainties.

\begin{figure}[htbp!]
  \centering
  \includegraphics[width=0.48\textwidth]{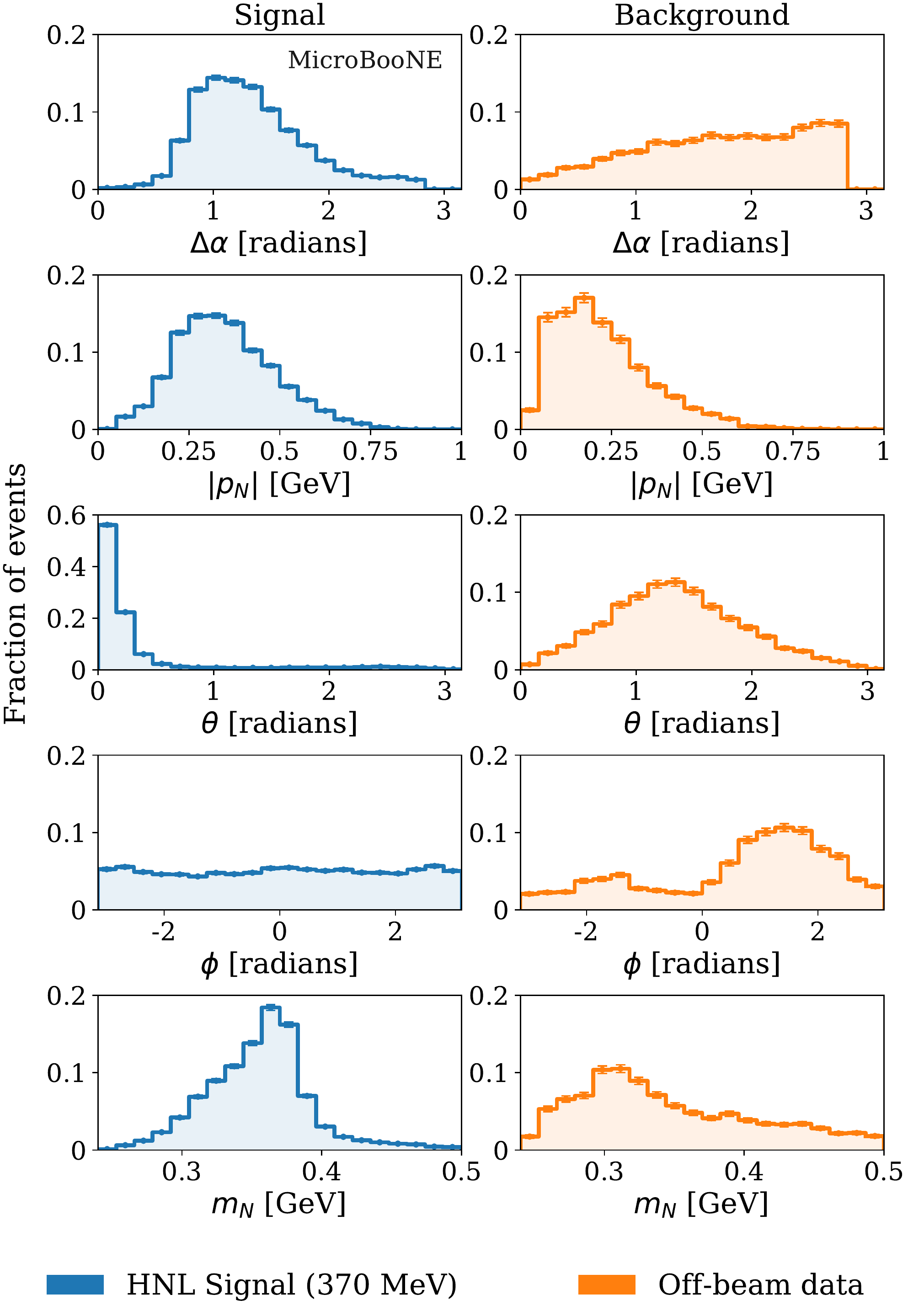}
  \caption{Kinematic variables used to train the BDT for the HNL candidates. The left column shows the distributions for HNL candidates with a mass of $370$~MeV, and the right column shows the distributions for the off-beam data. 
  The distributions are normalized to $1$.}
  \label{fig:bdtIsFine}
\end{figure}

\section{Signal Extraction}
\label{sec:signal_extraction}
We train a boosted decision tree (BDT) on a set of kinematic variables to discriminate between signal candidates and background using the XGBoost framework~\cite{Chen:2016:XST:2939672.2939785}. The following variables are used as input to the BDT:
\begin{itemize}
    \item The 3D opening angle $\Delta\alpha$ between the two tracks associated with the HNL decay;
    \item The momentum $\left| p_N \right|$ of the HNL candidate; 
    \item The polar angle $\theta$ of the HNL candidate;
    \item The azimuthal angle $\phi$ of the HNL candidate;
    \item The invariant mass $m_N$ of the $\mu\pi$ pair,
\end{itemize}
where $\theta$ and $\phi$ are defined in the MicroBooNE coordinate system.

Figure~\ref{fig:bdtIsFine} shows the distributions of these variables for signal HNL candidates with a mass
of $370$~MeV, and for the
off-beam data set with the selection applied. 
Since we observe no statistically significant difference in the kinematic distributions between the BNB and HNL triggered off-beam data sets, the higher statistics off-beam BNB data set is shown here.

\begin{figure*}[ht!]
  \centering
  \includegraphics[width=0.3\textwidth]{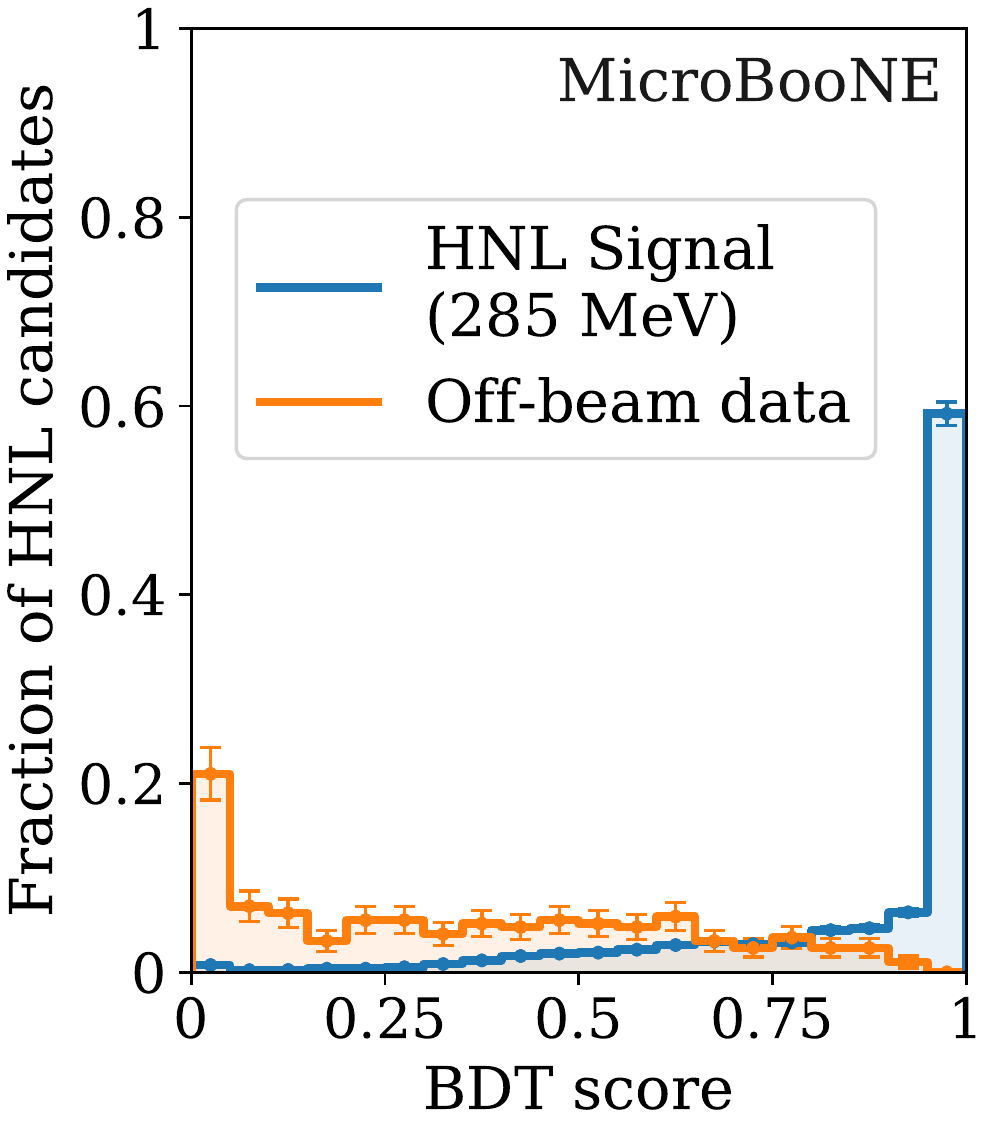}
  \centering
  \includegraphics[width=0.3\textwidth]{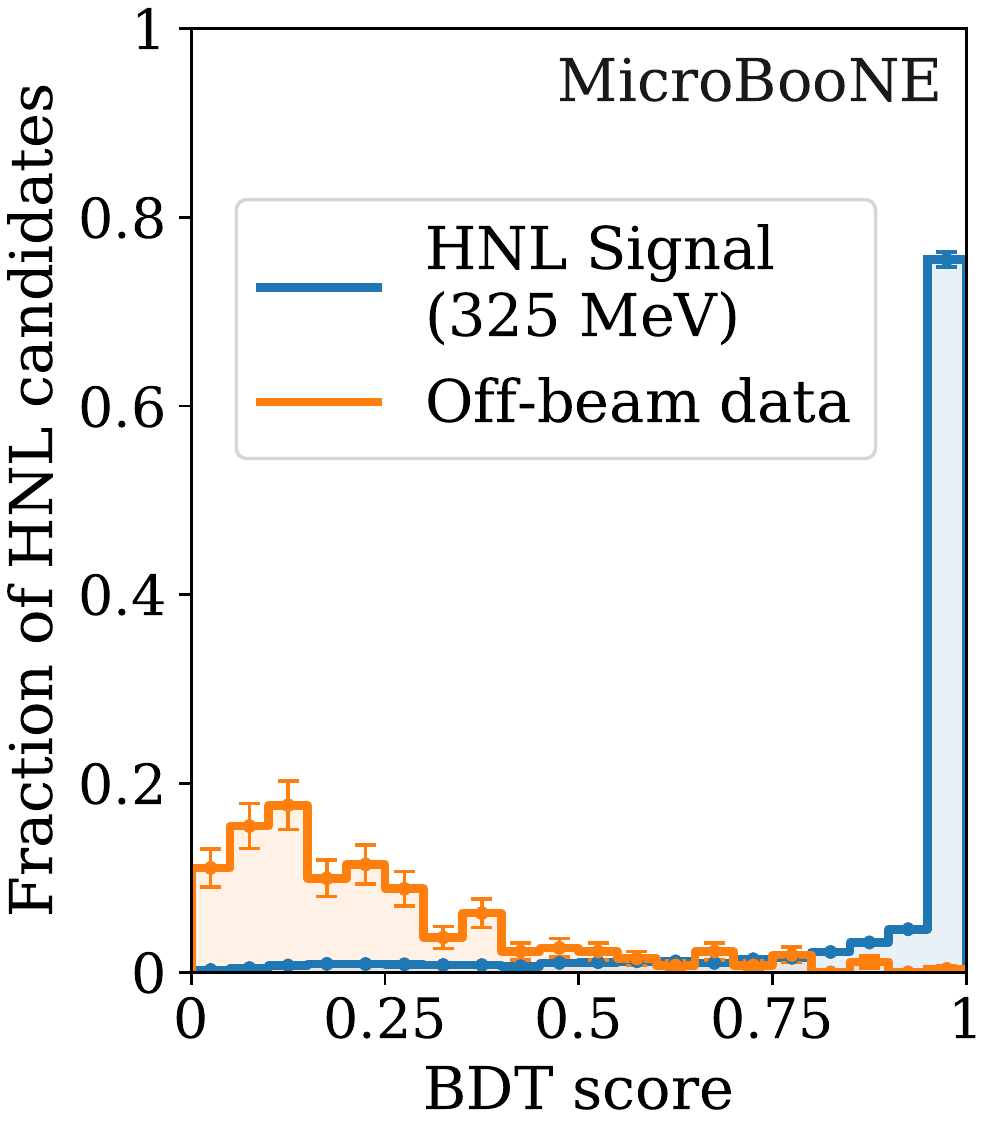}
  \includegraphics[width=0.3\textwidth]{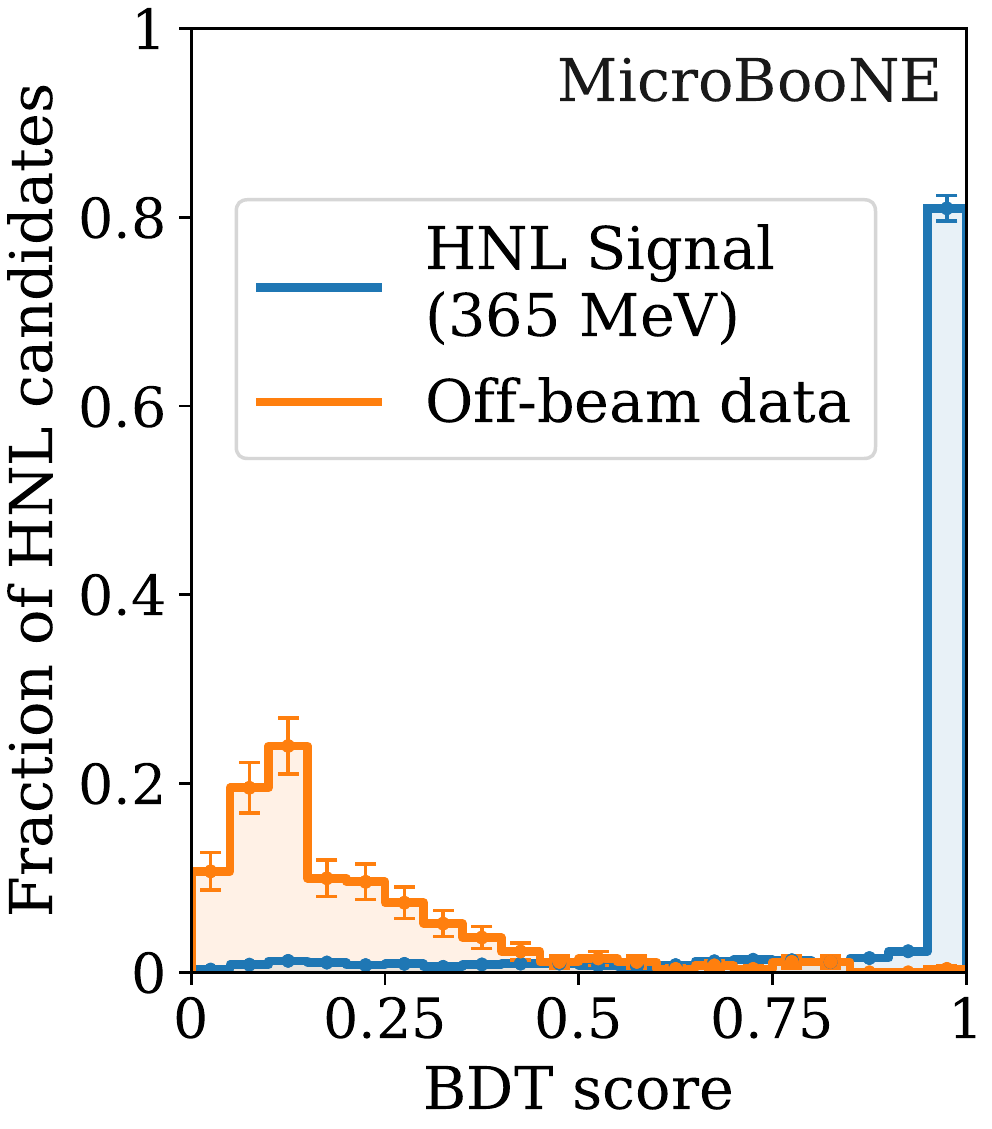}
  \caption{BDT score distribution 
  for three distinct BDTs trained with HNL masses of 285, 325, and 365 MeV, respectively, comparing simulated HNL signal and off-beam background data. The uncertainty on the samples is statistical.}
  \label{fig:testBDT1}
\end{figure*}

The distributions show good
separation between signal and background. Background is constituted mostly of ``broken" cosmic-ray tracks.
We train separate BDT models for ten different mass hypotheses in the range
$260$--$385$~MeV. 
The BDT score distributions for the signal and for the off-beam background shown in
Fig.~\ref{fig:testBDT1} demonstrate good separation between signal and background for different
HNL masses. The discrimination between signal and background improves with 
increased HNL mass. 

The distributions in Fig.~\ref{fig:bdtIsFine} are given for HNL candidates, where each event can contain more than one HNL candidate. Since the probability to observe more than one HNL candidate is negligible
for the mixing angles considered here, we retain only the candidate with the most signal-like BDT score
in each event.

\section{Control Samples}
\label{sec:control}
We use a statistically independent control sample to validate the determination of the reconstruction and selection efficiencies, and the BDT performance.
The control sample is the on-beam BNB data set, which is expected to contain SM neutrino interactions with
similar final-state topologies as HNL decays. 
In addition to applying the HNL selection to the control sample, 
we reject HNL candidates where one
of the tracks has an associated energy deposit consistent with the specific energy loss
expected for a proton. This gives a better representation of the HNL selection, since candidates with protons in the final state are not expected in the signal sample.

The trigger used to record the on-beam BNB sample selects events with any kind of interaction, i.e., 
the data sample will contain events with cosmic-rays only and no SM neutrino interaction.
We therefore
subtract the distributions obtained from an off-beam data sample containing cosmic-ray events to obtain distributions statistically representing a sample of mainly SM neutrino interactions. 

\begin{table}[htbp]
\centering
\caption{Numbers of HNL candidates remaining after the application of selection requirements to 
HNL candidates in the control samples, comparing simulation and data. The off-beam cosmic-ray data has
been rescaled and subtracted from the on-beam data to obtain the control sample. The uncertainty on the simulation is given by the statistical uncertainty on the generated number of MC events.}
\begin{tabular}{c|cc}
\hline\hline
                         & Control Sample & Simulation \\ \hline
Two-track Vertex         & 81,112 & 79,365          \\
Fiducial Volume          & 43,078  & 42,414         \\
Vertex-track Distance    & 32,120 & 32,471         \\
Minimum Number of Hits   & 31,939 & 32,228         \\
Flash Requirement        & 23,089 & 19,962          \\
Track Containment        & 6,344  & 6,021           \\
Kinematics (mass, angle) & 3,972  & 4,267$\pm$475   \\
\hline\hline
\end{tabular}
\label{tab:qualityCuts_numEvents}
\end{table}

\begin{figure*}[ht!]
  \centering
  \includegraphics[width=0.3\textwidth]{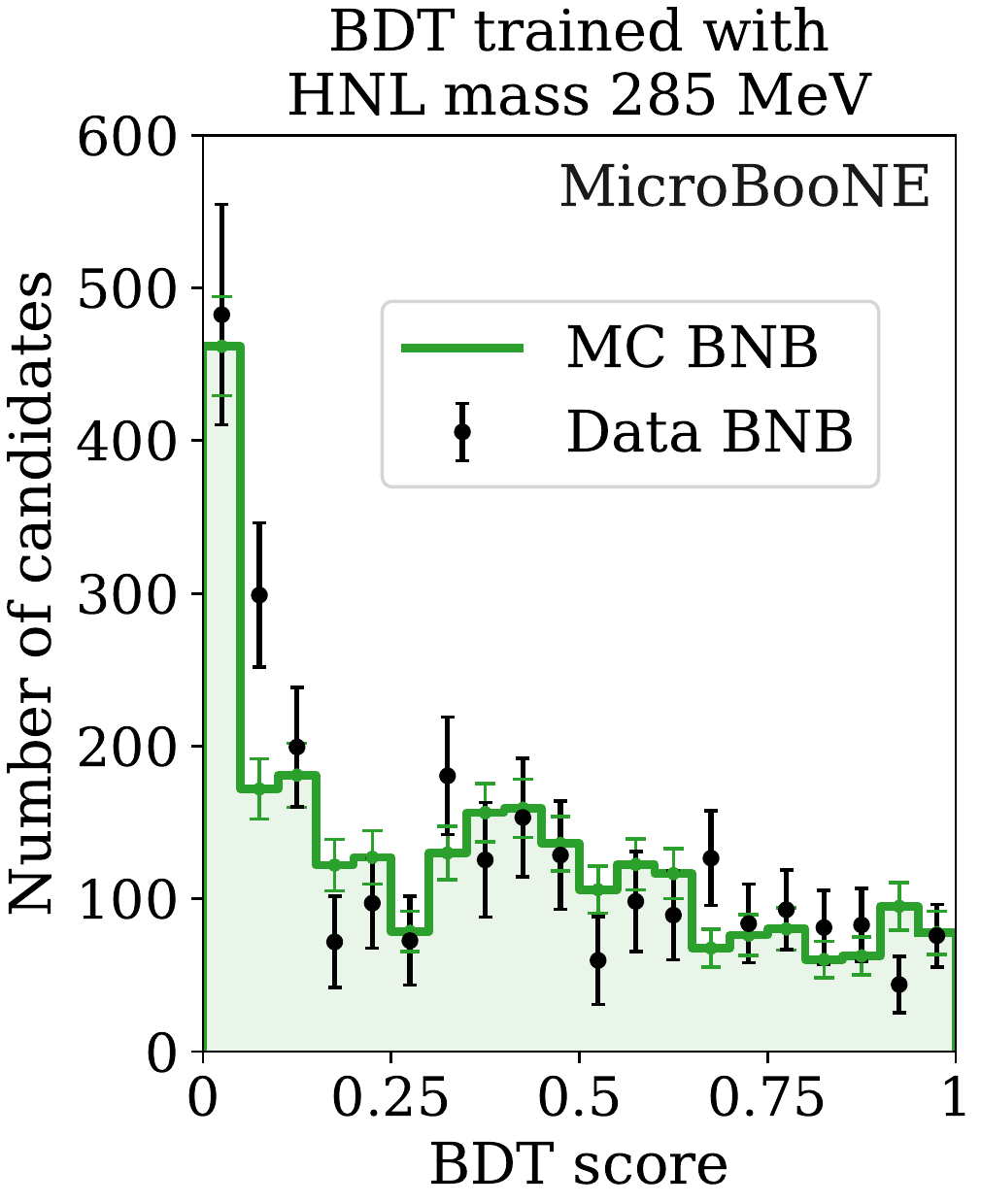}
  \includegraphics[width=0.3\textwidth]{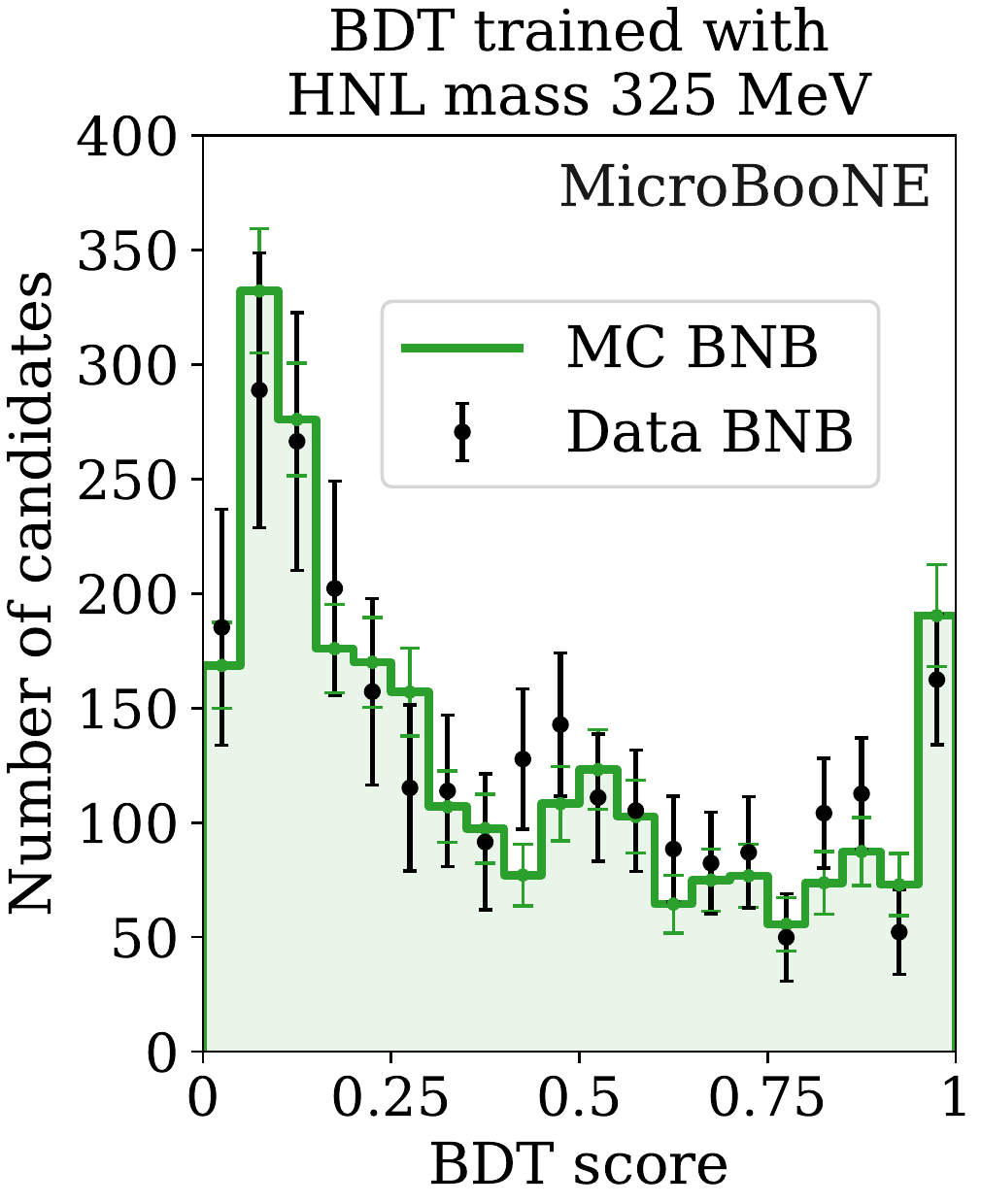}
  \includegraphics[width=0.3\textwidth]{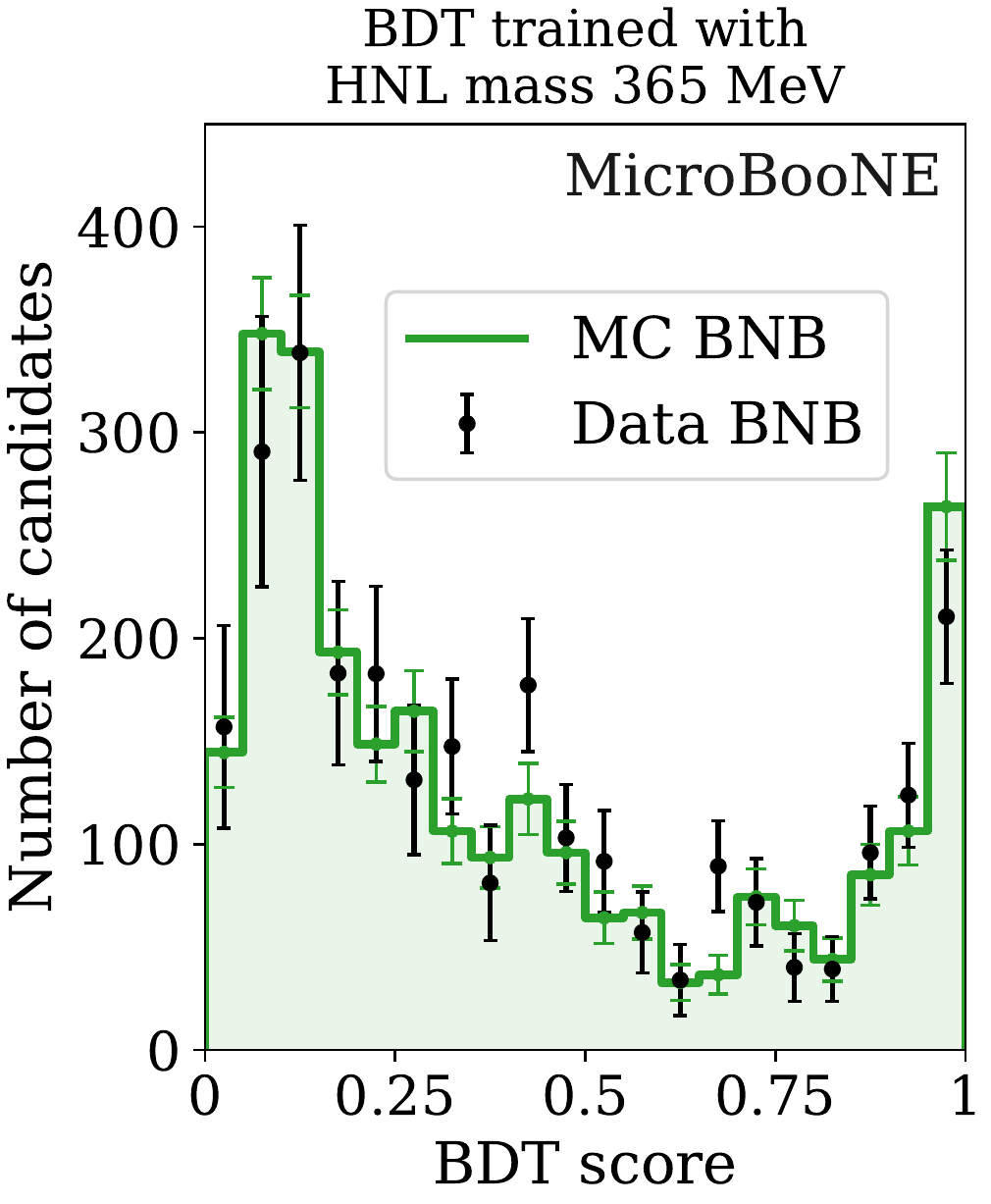}
  \caption{BDT score distributions for three distinct BDTs trained with HNL masses of 285, 325, and 365 MeV, respectively, of the MC BNB, and the on-beam BNB data subtracting the off-beam BNB control data samples. The uncertainty on the samples is statistical. Signal-like events are expected at higher BDT scores.}
  \label{fig:testBDT2}
\end{figure*}

The numbers of HNL candidates remaining after each preselection step are given in Table \ref{tab:qualityCuts_numEvents}
for the control sample and the Monte Carlo (MC) simulation of BNB SM neutrino events.
We then apply the same BDTs trained for
three HNL masses as used for the signal selection in the previous section
to the control samples after the selection to obtain the BDT score distributions in Fig.~\ref{fig:testBDT2}. 
The distributions for the MC simulation of BNB SM neutrino events are
normalized to the same number of POTs as the data.
We observe good agreement between data and simulation in terms of shape and normalization. 
For BDTs trained with higher HNL masses, signal-like events appear at higher BDT scores.
In the MC simulation, these events are mostly due to charged-current (CC) neutrino interactions.
Kinematics and track length in CC neutrino interactions become more similar
to the simulated kinematics for higher mass HNLs. This background is not present in the data samples
used for the HNL search.

\section{Systematic Uncertainties}
\label{sec:systematics}
Systematic uncertainties on the simulation of the HNL signal originate from the simulation of the HNL flux, 
the trigger efficiency,
and several calibration and reconstruction effects. To quantifty the systematic uncertainties, we
use a signal-enriched sample with a BDT score $>0.95$. The following uncertainties on the signal are considered:
\begin{itemize}
\item HNL flux
uncertainties are estimated by simultaneously varying all parameters used for the flux simulation~\cite{fluxnote}. 
The relevant parameters are related to the beamline simulation,
which includes variations of the horn current and the skin effect,
and to uncertainties on the kaon production cross section. 
The contributions to the total flux uncertainty from both types of uncertainty
are approximately equal.
For all HNL masses, the overall uncertainty 
on the flux is $8\%$. 
In addition, a constant $2\%$ uncertainty accounts for the POT counting performed with the beam toroid.

\item Systematic uncertainties on the trigger efficiency originate from the timing resolution
of the PMTs, which we use to define the HNL trigger window and vetoes of the BNB trigger window. Uncertainties caused by variations of the light yield are negligible. The trigger uncertainty is estimated to be in the range $(5-10)\%$ depending on the mass of the HNL signal.

\item Dynamically induced charge (DIC) 
refers to the charge induced on the wires beyond the 
wire closest to the ionization trail~\cite{Adams:2018dra,Adams:2018gbi}. 
This effect impacts the algorithm that determines the deposited charge, especially
on induction planes, as well as the pattern recognition. 
To estimate the impact of DIC effects, we compare
samples where DIC is simulated in the region up to the adjacent $20$ wires ($10$ on each side) with a MC simulation that does not model DIC. 
The resulting uncertainty on the
normalization of the signal sample is $\approx 10\%$.

\item Distortions caused by the SCE are corrected by time-averaged calibration
factors obtained from data. We estimate the uncertainties originating from the SCE correction by comparing the reconstruction efficiency in the MC samples with and without the simulation of the SCE. The corresponding uncertainty is found to be negligible ($<1\%$).

\item The uncertainties from the remaining detector effects (such as recombination, attenuation and diffusion) are determined by comparing HNL MC samples that only differ in the simulation of the detector response. 
The response is either obtained from data after detector calibration or from simulation.  The resulting uncertainties due to the detector response are estimated to be small ($<1\%$).
\end{itemize}
Uncertainties from nuclear interaction modelling are negligible, since the main background source are cosmic-ray muons. For Dirac HNLs, we consider only $\mu^{-}\pi^{+}$ decays and not the sum of the charge combinations.
The difference in efficiency between the charge combinations is $\approx (2-3)\%$, which is small compared to the total systematic uncertainty. This difference is thus neglected.
\begin{table}[ht!]
\centering
\caption{Systematic uncertainties in the signal sample with BDT score $>0.95$ for an HNL mass value of $325$~MeV.}
\begin{tabular}{lc}
\hline\hline
Source & Uncertainty \\ \hline
HNL Flux & $8 \%$ \\
POT & $2\%$ \\
Trigger & $8\%$ \\
Dynamically Induced Charge & $10 \%$ \\
Space Charge Effect & $0.3 \%$ \\
Detector Response & $0.4 \%$ \\
\hline
Total & $15 \%$ \\
\hline\hline
\end{tabular}
\label{tab:syst}
\end{table}

\begin{figure*}[hbt!]
  \centering
  \includegraphics[width=0.3\textwidth]{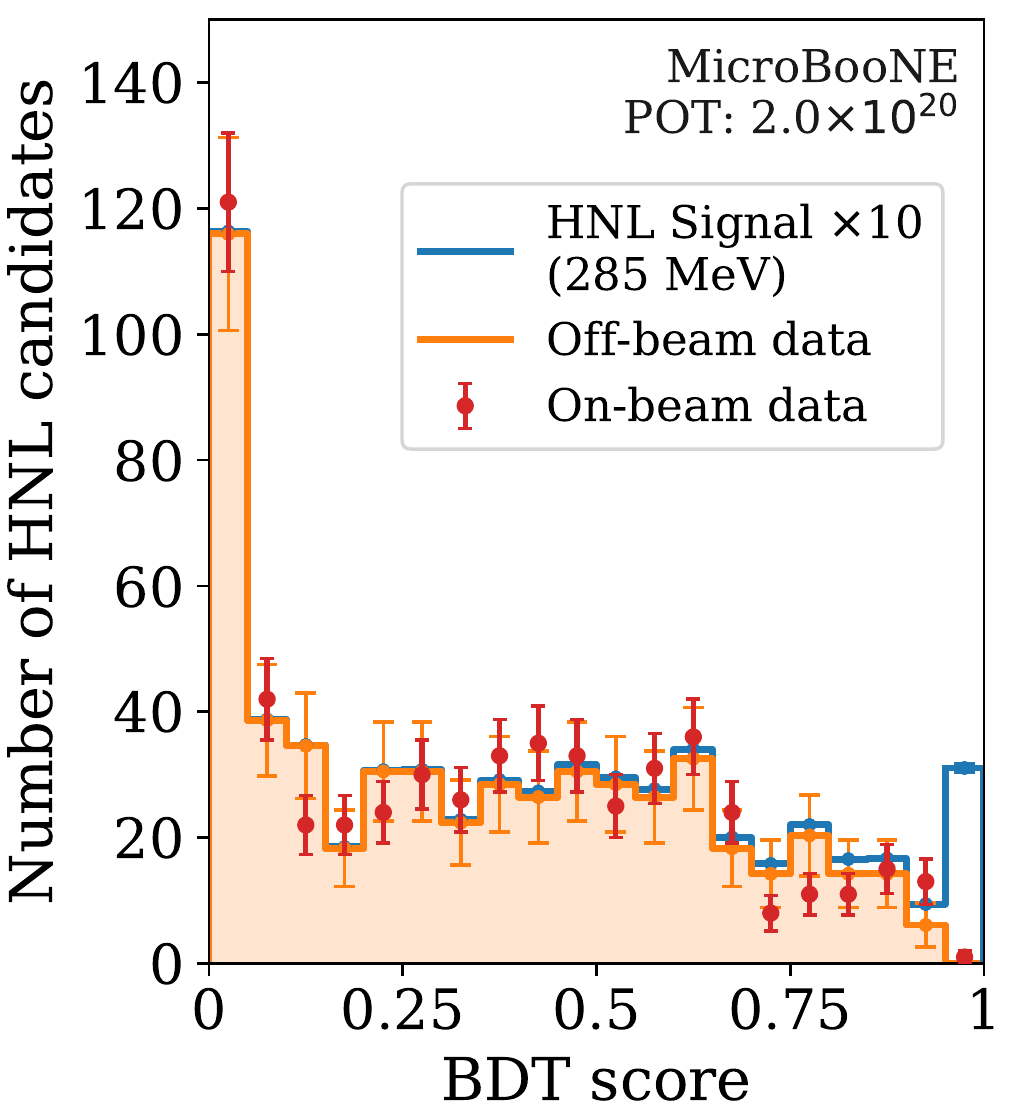}
  \includegraphics[width=0.3\textwidth]{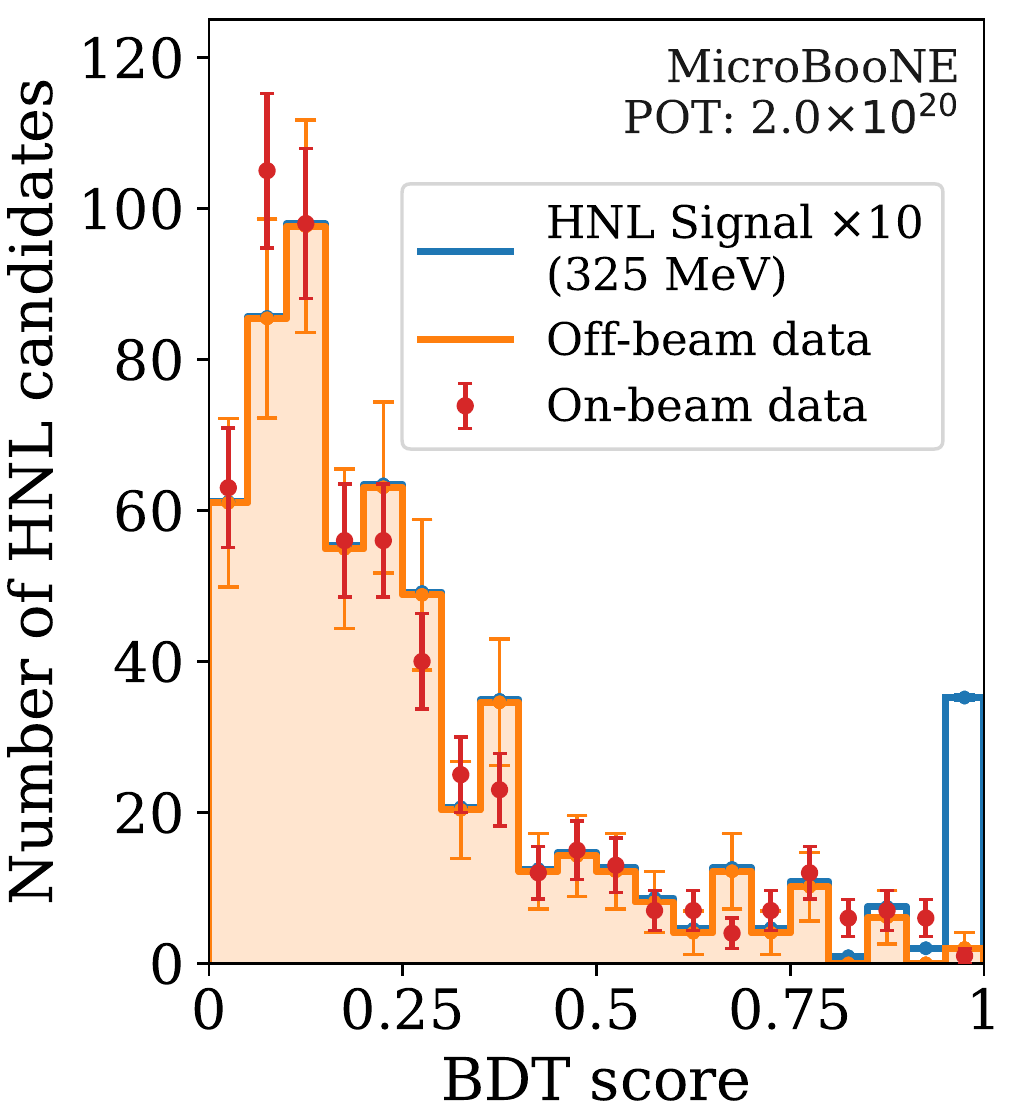}
  \includegraphics[width=0.3\textwidth]{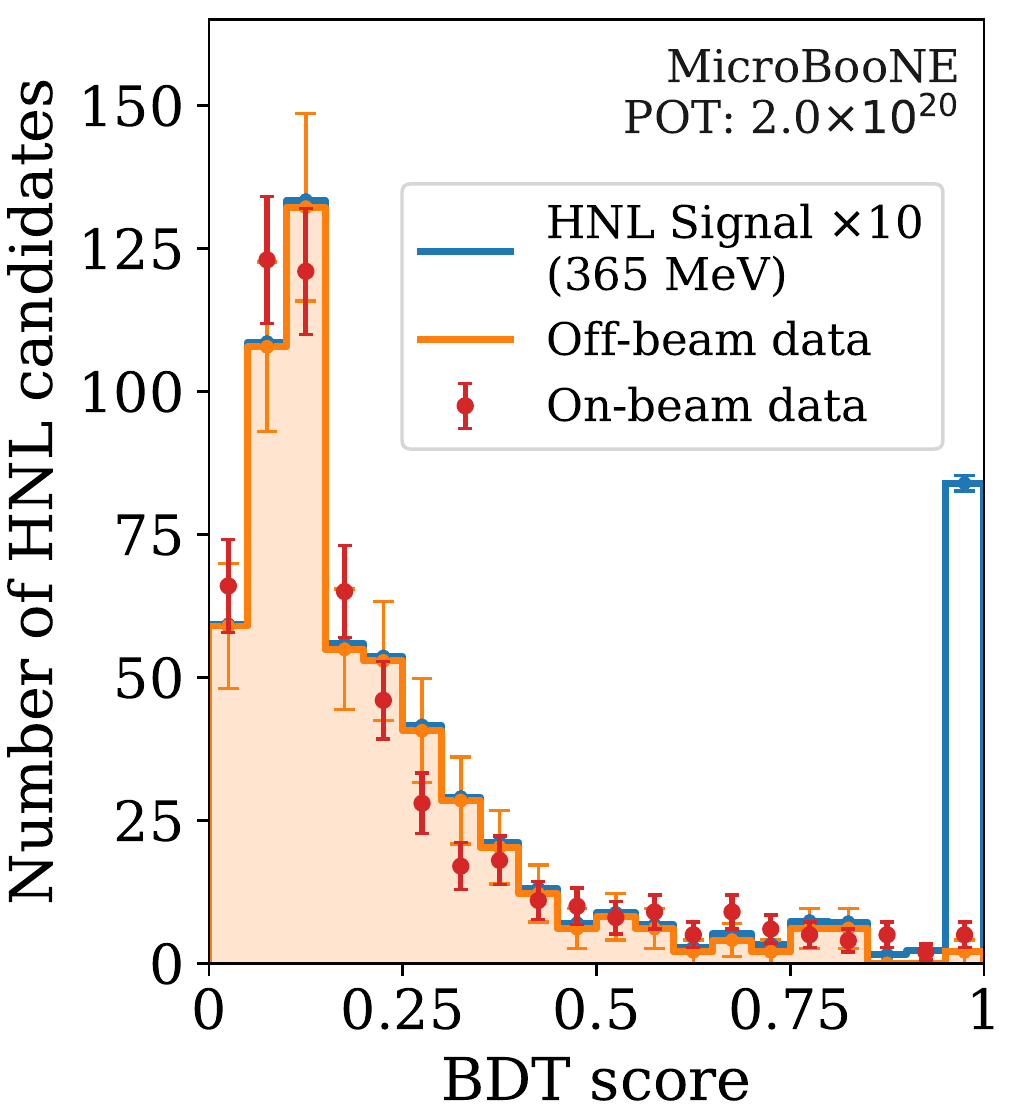}
  \caption{BDT score distribution for the on-beam HNL and background off-beam HNL data samples. 
  The signal distribution is shown as a stacked histogram added to the background (off-beam data), with the normalization fixed at
  the $90\%$ CL level limit, multiplied by a factor of 10.  The statistical uncertainties for the data samples
  are shown separately.}
  \label{fig:testBDT3}
\end{figure*}
\vspace{-1mm}
The contribution of the systematic uncertainties on the
signal efficiency in the signal-enriched sample
are summarized in Table~\ref{tab:syst} for a HNL mass value of 325 MeV. The systematic uncertainties grow linearly with HNL mass, from 10\% to 18\% in the mass range $260$--$385$~MeV.
Uncertainties on the background estimation, which is derived from data, is dominated by the statistical
fluctuations of the off-beam HNL data sample. 

\vspace{-3.8mm}
\section{Results and Discussion}
\label{sec:results}
The BDT score distributions for signal, background expectation (off-beam HNL data) and data are shown in
Fig.~\ref{fig:testBDT3}. Signal and background are well separated, and no data excess is observed in the signal region
with high BDT score. We therefore proceed to set limits on the HNL production rate as a function of mass.
\begin{table}[htbp!]
\centering
\caption{  
Number of events with higher BDT score of $>0.95$ and BDT score 
in the range $0.5$ to $0.95$ for an HNL signal with
$\umusq= 1.4 \times 10^{-7}$, for the expected background, and for the on-beam HNL data. Systematic uncertainties are given for the signal and statistical uncertainties for the background. The $68\%$~CL Poisson interval is used for bins with zero expected background events.} 
\begin{tabular}{c|ccc|ccc}
\hline\hline
Mass & \multicolumn{3}{c|}{BDT Score $>0.95$} & \multicolumn{3}{c}{BDT Score $0.5$--$0.95$}\\
(MeV) & HNL & Bkg. & Data &  HNL & Bkg. & Data  \\ \hline
260 & $0.21  \pm 0.03$ & $< 3.7$ & 1 & $0.43 \pm 0.06$  & $169 \pm 19\phantom{0}$ & $170$\\
265 & $0.42  \pm 0.06$ & $2 \pm 2$ & 1 & $0.6 \pm 0.1$  & $185 \pm 19\phantom{0}$ & $205$ \\
285 & $1.6 \pm 0.3$    & $<3.7$ & 3 & $0.8 \pm 0.1$  & $175 \pm 19\phantom{0}$ & $174$\\
300 & $\phantom{0}2 \pm 0.3$      & $2 \pm 2$ & 1 & $1.0 \pm 0.2$  & $126 \pm 16\phantom{0}$ & $121$ \\
305 & $\phantom{0}4 \pm 0.6$      & $2 \pm 2$ & 4 & $0.8 \pm 0.1$  & $61 \pm 11$ & $80$\\
325 & $6  \pm 1$       & $2 \pm 2$ & 0 & $1.6 \pm 0.3$  & $57 \pm 11$ & $69$\\
345 & $12 \pm 2\phantom{0}$       & $2 \pm 2$ & 4 & $\phantom{0}2 \pm 0.3$    & $59 \pm 11$ & $69$ \\
365 & $20 \pm 3\phantom{0}$       & $2 \pm 2$ & 5 & $\phantom{0}2 \pm 0.3$    & $35 \pm 8\phantom{0}$ & $53$ \\
370 & $24 \pm 4\phantom{0}$       & $2 \pm 2$ & 4  & $\phantom{0}4 \pm 0.6$    & $37 \pm 9\phantom{0}$ & $47$\\
385 & $36 \pm 6\phantom{0}$       & $< 3.7$ & 4 & $\phantom{0}4 \pm 0.6$    & $20 \pm 6\phantom{0}$ & $28$  \\
\hline\hline
\end{tabular}
\label{tab:finalCount_b0}
\end{table}

The limits are determined using the modified frequentist CL$_s$ method~\cite{Junk:1999kv,Read:2002hq,Fisher:2006zz}.
We calculate a log-likelihood ratio (LLR) test statistic using Poisson probabilities for estimated
background events, signal yields, and the observed number of events for different HNL mass hypotheses.
The confidence levels are derived by integrating the LLR distribution in pseudo-experiments using both
the signal-plus-background (CL$_{s+b}$) and the background-only hypotheses (CL$_b$). The excluded signal rate 
is defined by the signal strength for which the confidence level for signal, CL$_s$ = CL$_{s+b}$/CL$_b$, equals 0.1.

\begin{figure*}[htbp!]
    \centering
    \includegraphics[width=1.4\columnwidth]{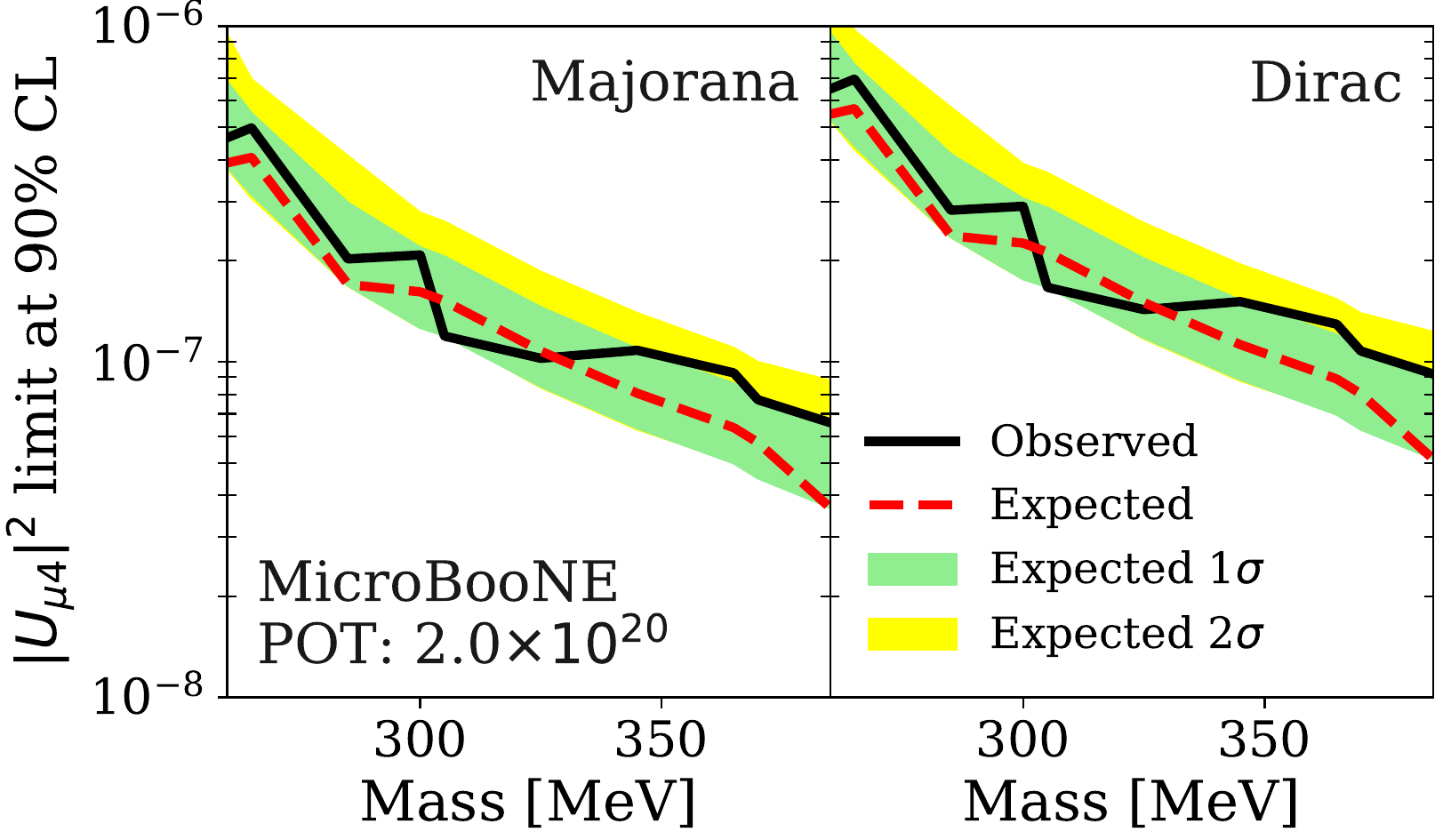}
    \caption{Limits on $\umusq$ at the $90\%$ confidence level as function of mass for a Majorana and Dirac HNL decaying into $\mu\pi$ pairs. The observed limit is compared
    to the median expected limit with the $1$ and $2$ standard deviations ($\sigma$) bands.}
    \label{fig:limit}
\end{figure*}

Systematic uncertainties on both background and signal are taken into account using Gaussian priors.
The total systematic uncertainty of approximately $15\%$ on the signal is found to have a negligible impact on the sensitivity,
since the uncertainty is dominated by the statistics of the background sample.

To calculate limits, we split the distribution into a signal-enriched and signal-depleted region
with BDT scores $> 0.95$ and $0.5<\text{BDT score}<0.95$, respectively.
In Table~\ref{tab:finalCount_b0}, we compare the expected number of
background events to the data for the different HNL mass hypotheses, as well as for the signal-enriched and signal-depleted samples.
The expected number of HNL signal events is calculated assuming $\umusq= 1.4 \times 10^{-7}$.

The observed upper limits at the $90\%$ confidence level as a function of mass
are presented in Fig.~\ref{fig:limit} and Table~\ref{tab:brazilTable} for Majorana HNLs, together with the median expected limit and the $1$ and $2$ standard deviation bands on the median expected limit. The bands are asymmetric since Poisson statistics are used where the expected number of events is small.
The observed and expected limit agree within $1$ standard deviation over the entire mass range. The decay rates for Dirac HNLs are a factor of $2$ smaller, and we observe no significant difference between the efficiencies
to observe Majorana and Dirac HNLs. Limits for the Dirac case are  therefore derived by multiplying the values in Table~\ref{tab:brazilTable} by a factor of $\sqrt{2}$.

\begin{table}[ht!]
    \vspace{-4mm}
\centering
\caption{Limits on $\umusq$ at the $90\%$ confidence level for Majorana HNLs decaying into $\mu\pi$ pairs, multiplied by a factor of $10^{7}$.
The Majorana HNL limits are multiplied by a factor of $\sqrt{2}$ to
obtain the Dirac HNL limits.}
\begin{tabular}{c|ccc}
\hline\hline
Mass (MeV) & Obs. & Median Exp.& $1\sigma$ band \\\hline
260 & 4.65 & 3.92  & $3.81$--$6.78$ \\
265 & 4.98 & 4.06  & $3.16$--$5.49$ \\
285 & 2.03 & 1.70  & $1.69$--$2.97$ \\
300 & 2.08 & 1.62  & $1.27$--$2.19$ \\
305 & 1.19 & 1.52  & $1.20$--$2.05$ \\
325 & 1.02 & 1.08  & $0.84$--$1.41$ \\
345 & 1.08 & 0.80  & $0.63$--$1.05$ \\
365 & 0.92 & 0.63 & $0.50$--$0.86$ \\
370 & 0.77 & 0.57 & $0.45$--$0.77$ \\
385 & 0.65 & 0.36 & $0.36$--$0.63$ \\
\hline\hline
\end{tabular}
\label{tab:brazilTable}
\end{table}

The results are of similar or better sensitivity as those obtained by the
NA62~\cite{CortinaGil:2017mqf} and NuTeV~\cite{Vaitaitis:1999wq} Collaborations for
the same mixing parameter in the overlapping HNL mass range.
The E949 Collaboration~\cite{Artamonov:2014urb} sets limits in a lower mass range between
$175$ and $300$~MeV by measuring $K^+$ meson decays at rest.
The PS191~\cite{Bernardi:1985ny,Bernardi:1987ek} and T2K~\cite{Abe:2019kgx}
Collaborations place more stringent limits in the HNL mass range between $260$ and $360$~MeV, but their location at an off-axis angle of $2$~degrees restricts the sensitivity to slightly lower masses. 
The MicroBooNE detector is located on-axis, allowing it 
to set the most constraining limits in the mass range up to $385$~MeV.

\vspace{-3.8mm}
\section{Conclusion}
\label{sec:summary}

We present the first search for HNLs in a liquid-argon TPC 
using data recorded with the MicroBooNE detector in 
$2017$--$2018$ with a novel trigger
recording events that arrive at the MicroBooNE detector after the BNB beam spill. The data 
correspond to $2.0 \times 10^{20}$ POT.
We assume that the HNLs are produced in kaon decays and 
decay exclusively into a muon pion final state. 
We obtain constraints on the element $\umusq$ of the extended PMNS mixing matrix of
$\umusq<(4.7$--$0.7)\times 10^{-7}$ for Majorana HNLs and $\umusq<(6.6$--$0.9)\times 10^{-7}$ for Dirac HNLs with masses between $260$ and $385$~MeV and 
assuming $\lvert U_{e 4}\rvert^2 = \lvert U_{\tau 4}\rvert^2 = 0$. 
\vspace{-3.8mm}
\section*{Acknowledgements}
\label{sec:acknowledgements}
This document was prepared by the MicroBooNE Collaboration using the resources of the Fermi National
Accelerator Laboratory (Fermilab), a U.S. Department of Energy, Office of Science, HEP User Facility. Fermilab is managed 
by Fermi Research Alliance, LLC (FRA), acting under Contract No. DE-AC02-07CH11359. MicroBooNE is supported by the 
following: the U.S. Department of Energy, Office of Science, Offices of High Energy Physics and Nuclear Physics; the U.S. 
National Science Foundation; the Swiss National Science Foundation; the Science and Technology Facilities Council (STFC), 
part of the United Kingdom Research and Innovation; and The Royal Society (United Kingdom). Additional support for the 
laser calibration system and cosmic ray tagger was provided by the Albert Einstein Center for Fundamental Physics, Bern, 
Switzerland.
\bibliography{HNL}

\end{document}